\documentclass[12pt]{article}
\usepackage{epsfig}
\begin{document}

\title
{Fractal Analysis for Social Systems \\}

\author{\bf C.M. Arizmendi
\\
\rm \normalsize Departamento de F\'{\i}sica, Facultad de Ingenier\'{\i}a\\
\rm \normalsize Universidad Nacional de Mar del Plata\\
\rm \normalsize Av. J.B. Justo 4302\\
\rm \normalsize 7600 Mar del Plata\\
\rm \normalsize Argentina}
\date{}
\maketitle

\begin{abstract}
This is a brief introduction to fractals, multifractals and wavelets  in an 
accessible way, in order that the founding ideas of those strange and 
intriguing newcomers to science as fractals may be communicated to a wider 
public. Fractals are the geometry of the wildness of nature, where the 
euclidian geometry fails. The structures of nonlinear dynamics associated with 
chaos are fractal. Fractals may also be used as the geometry of social systems.
 Wavelets are introduced as a tool for fractal analysis. As an example of its 
 application on a social system, we use wavelet fractal analysis to compare  
 electrical power 
 demand of two different  places, a touristic city and a whole country. 
 
 \end{abstract}
\newpage
\section{Introduction}
People have been trying to make life structured and organized throughout
recorded time (and probably before). But, nature is not orderly and  the social world is not orderly. A good example are the capital markets. Models have been created to explain them. These
models are, of necessity, simplifications of reality. By making a few
simplifying assumption about the way investors behave, an entire analytic
framework has been created to help us understand the markets. The models have
not worked well. Studies of economic forecasts \cite{mcnees,forbes} show that
economists have made serious forecasting errors at every major turning points
since the early 1970s, when the studies began. Included in the group studied was
Townsend-Greenspan, run by Fed Chairman Alan Greenspan. Forecasters tend to be
out as a group at these turning points. What went wrong?

Econometric analysis assumes that, without outside influences everything
balances out. Supply equals demand. If exogenous factors perturb the system,
taking it away from equilibrium, the system reacts inmediately reverting to
equilibrium in a linear fashion. But a free-market economy is an evolving
structure with emotional forces, such as greed and fear, which cause the economy
to develop  ``far from equilibrium" conditions sometimes.

An ``efficient market" is one in which assets are fairly priced and neither
buyers nor sellers have advantage. However, new financial instruments with low
interest eventually die, even if they are fairly priced. Any trader will confirm
that a healthy market is one with volatility, but not necessarily fair price. We
may say that a healthy economy and a healthy market {\sl do not} tend to
equilibrium but are, instead, far from equilibrium and equilibrium theories are
likely to produce dubious results.

Another problem is that with the econometric view of the world, the markets and
the economy have no memory, or only limited memory of the past. As an example,
let us say that interest rates $r$ depend solely on the current rate of
inflation $i$ and the money supply $s$. A simple model would be:
\begin{equation}
r= ai+bs.
\end{equation}

If the coefficients $a$ and $b$ are fixed, then $r$ depends on current levels of
$i$ and $s$. It does not matter whether $i$ and $s$ are rising or falling. What
is missing, is the feedback effect produced by the fact that, in human decision
making, the expectations of the future are influenced by recent experiences.
Feedback systems are characterized by long-term correlations and trends.\\
These characteristics - far from equilibrium conditions and feedback mechanisms
- are symptomatic of nonlinear dynamic systems. Nonlinear differential - or
difference - equations are complex and have multiple, messy solutions. Life is
messy, there are many possibilities.

Let's illustrate with a simple, nonlinear model related with a social system.
Suppose a new TV program with audience (normalized) $R_t$. The audience rises at a rate $a$. Considering only this effect, the audience  
would increase as:
\begin{equation}
R_{t+1} = a R_t.
\end{equation}

But there are spectators that may see the TV show and do not like it. 
We may suppose that this effect cause a reduction of audience at $aR_t^2$.
 The evolution of the audience rating would then be:
\begin{equation}
R_{t+1} = a R_t(1-R_t).
\end{equation}

Although this is a simple model, it explains that at low levels of audience 
growing $a$ ($a<1$), the audience goes to zero (and the TV managers get rid 
of the show) and, at higher levels of audience growing, the audience will 
converge to a steady  value.

\begin{figure}[t]
\begin{center}
\psfig{figure=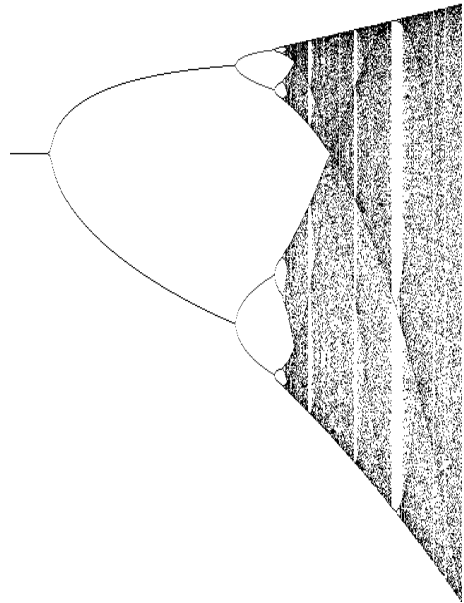,height=8cm,width=6cm}
\caption{\tt Bifurcation diagram of the logistic equation }
\end{center}
\end{figure} 
Let us see what happens for $a>1$.
Suppose a growth rate of $a=2$, and $R_0=0.1$. By iterating equation (3), a steady audience  of $0.5$ is reached. You can try with your personal computer, and a spreadsheet, copying equation (3) down for 200 cells approximately. Repeating this experiment for different initial values $R_0$ the same final audience of $0.5$ is obtained. If the growth rate is increased to $a=2.5$, two possible final audiences appear, and the system oscillates between them. As the growth rate continues to rise, each possible audience bifurcates and 4, 8, and 16 final audiences appear. Finally, at $a=3.5699456..$, the system displays an infinite number of possible final values, fluctuating between them in a chaotic way. The changes from one audience to the next one seem absolutely random, as though blown about by environmental noise. Yet in the middle of this complexity, stable cycles return. Even though the parameter is rising, meaning that the nonlinearity is driving the system harder, a window will suddenly appear with a regular period: an odd period, like $3$ or $7$. Then the period-doubling bifurcations  begin all over at a faster rate, rapidly passing through cycles of 3, 6, 12,... or 7, 14, 28,..., and then breaking off again to chaos. The final audience value behavior with increasing growth rate is shown in Fig. 1. 

Equation (3) is the famous ``Logistic Equation'', which was formulated in 1845 by Pierre Verhulst to model the growth of populations limited by finite resources. Its strange behavior discussed before was discovered by Robert May \cite{May} in 1976 and it was later used by Mitchell Feigenbaum for his ground breaking work on the universality of the period-doubling route to chaos \cite{Feigen,hein}.

This  model is, obviously, too simplified, because, for instance, it assumes that the audience decreasing is directly related to the growth rate $a$. However, it shows us the most important characteristics of nonlinear dynamic systems.

The first one is the {\it extremely sensitive dependence on initial conditions} which is the signature of chaos: A slight change in $R_t$ in the chaotic region will result in a completely different audience after $n$ steps. Around 1960, Edward Lorenz discovered (by accident) this characteristic in the models used for numerical weather forecasting; and it was he who coined the term ``butterfly effect''. He worked with twelve equations to simulate the atmosphere, by solving them with an MIT computer. In order to examine some of the results in more detail, he used a small computer that he had in his office to introduce intermediate conditions which the big computer had printed out as new initial conditions to start a new computation. But the solution that came out was completely different from that obtained with the big computer. Lorenz found that the reason was that the numbers that he used as new initial conditions were not the same as the original ones, they had been rounded off from six decimal places to three and the difference had amplified until being as big as the signal itself.

 You can repeat easily the Lorenz experiment with the Logistic equation (3), taking $a=4$, using a computer or even a pocket calculator by making two series of iterations with the same starting point. After $10$ iterations, in one of the series the output is truncated to three decimal places and taken as input for the following iteration. Soon afterwards (around $10$ steps) the outputs will be completely different. Another nice examples of this kind of experiments with different calculators and with different implementations of the same quadratic law on the same calculator may be found in \cite{hein}.

The other important characteristic that may be appreciated in Fig. 1 is that it is a {\it fractal}. In the windows of stability, inside each figure is a smaller figure, identical to the larger figure. Enlarging the smaller figure, another window of stability may be found, where another smaller version of the main figure appears. At smaller and smaller scales happens the same. This is called the {\it self similar} property, which is characteristic of {\it fractals}.

 \section{Fractals}

The father of fractals, Benoit Mandelbrot, begun thinking in them studying the distribution of large and small incomes in economy in the $60$'s. He noticed that an economist's article of faith seemed to be wrong. It was the conviction that small, transient changes had nothing in common with large, long term changes. Instead of separating tiny changes from grand ones, his picture bound them together. He found patterns across every scale. Each particular change was random and unpredictable. But the sequence of changes was independent of scale: curves of daily changes and monthly changes matched perfectly.
\begin{figure}[t]
\begin{center}
\psfig{figure=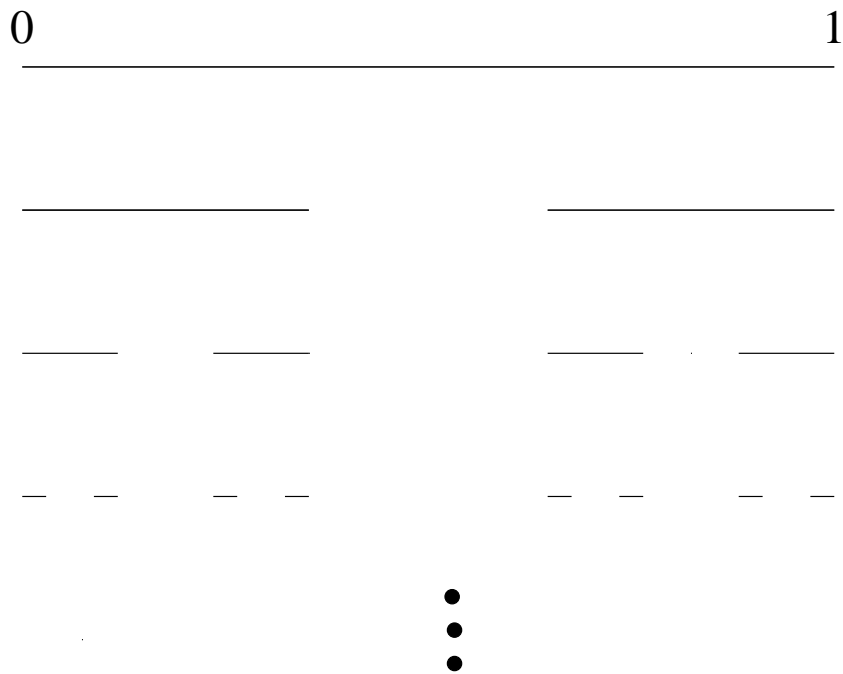}
\caption{\tt Initial steps of the construction of the Cantor Set}
\end{center}
\end{figure}
Mandelbrot worked in IBM, and after his study in economy, he came upon the problem of noise in telephone lines used to transmit information between computers. The transmission noise was well known to come in clusters. Periods of errorless communication would be followed by periods of errors. Mandelbrot provided a way of describing the distribution of errors that predicted exactly the pattern of errors observed. His description worked by making deeper and deeper separations between periods with errors and periods without errors. But within periods of errors (no matter how short) some periods completely clean will be found. 

Mandelbrot was duplicating the Cantor set, created by the $19^{th}$ century mathematician Georg Cantor. To make a Cantor set (see Fig. 2), you start with the line segment from $0$ to $1$. Then you remove the middle third. That leaves two segments, and you remove the middle third from each. That leaves four segments, and you remove the middle third from each - and so on to infinity. The Cantor set is the strange ``dust'' of points that remains. They are  arranged in clusters, infinitely many yet infinitely sparse and with total length $0$.

In the {\it fractal way} of looking nature roughness and asymmetry are not just 
accidents on the classic and smooth shapes of Euclidian geometry. Mandelbrot 
has said that ``mountains are not cones and clouds are not spheres''. Fractals 
have been named the {\sl geometry of nature} because they can be found 
everywhere in nature:  mammalian lungs,  trees, and  coastlines, to name just a few. 

An english scientist, Lewis Richardson around 1920 checked encyclopedias  in Spain, Portugal, Belgium and the Netherlands and discovered discrepancies of twenty percent in the lengths of their common frontiers. In \cite{hein} the authors measured the coast of Britain on a geographical map with different compass settings by counting the number of steps along the coast with each setting. The smaller the setting they used, the longer the length of the coastline they obtained. If this experiment is done to measure the perimeter of a circle, (or any other euclidean shape), the length obtained converges with smaller compass settings. In his famous book \cite{Mandel} Mandelbrot states that the length of a coastline can never be actually measured, because it depends on the length of the ruler we use.

Since length is not a valid way to compare coastlines, Mandelbrot proposes fractal dimension to measure the degree of roughness or irregularity of coastlines and other rough objects. The idea is that the degree of irregularity remains constant over different scales.

\begin{figure}[t]
\begin{center}
\psfig{figure=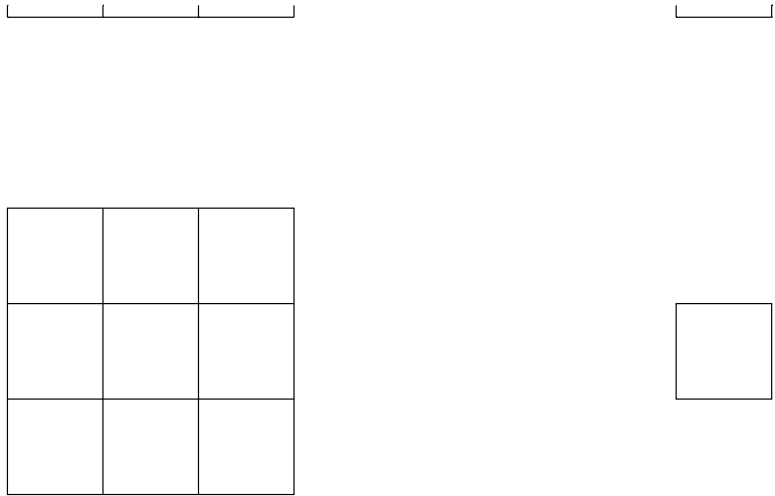}
\caption{\tt Self Similarity of Line and Square}
\end{center}
\end{figure}

We have previously discussed that fractals are self-similar, but a segment, or a square can be divided into small copies  (see Fig. 3). These structures, although self-similar are not fractals. There is a relation between the reduction or scaling factor $s$ and the number of scaled down pieces $N$ into which the  structure is divided. 
\begin{equation}
N=\frac{1}{s^D},
\end{equation}
where $D=1$ for the line and $D=2$ for the square, which agree exactly with the known (topological) dimensions of the segment and the square. It may be easily seen that in the Cantor set, $N$ scales as $2^{step}$ and $s=1/3$. Then $D=\frac{log(2)}{log(3)}$ for the Cantor set. $D$ is called the {\it self similarity dimension}. It is a special form of Mandelbrot's fractal dimension.

The most popular  version of Mandelbrot's fractal dimension  is the {\it box-counting dimension}, which is a concept related to  the self-similarity dimension and it is used in cases such as the coastlines where there is no exact self-similarity. The recipe of box-counting dimension calculation is to put the structure onto a regular mesh with mesh size $\epsilon$, and simply count the number of grid boxes which contain some of the structure. This gives a number $N$ which, of course will depend on the size $\epsilon$ of the mesh. Then change $\epsilon$ to progressively smaller sizes counting the corresponding $N(\epsilon)$. The scaling relation
linking $N(\epsilon)$ and the box-counting dimension $D_b$ is 
\begin{equation}
N(\epsilon) \sim \epsilon^{-D_b}.
\end{equation} 
Next make a log-log diagram and try to fit a straight line to the plotted points and measure its slope $D_b$. The box-counting dimension and the self-similarity dimension give the same numbers in many cases, as, for example, the Cantor set.

Fractal sets show self similarity with respect to space. Fractal time series have statistical self similarity with respect to time. In the social and economic fields, time series are very common. In \cite{pete} a simple way to demonstrate self-similarity in a time series of stock returns is devised by asking the reader to guess which graph corresponds to daily, weeekly  and monthly returns between three different graphs with no scale on the axes. 

An important statistics used to characterize time series is the {\it Hurst exponent} \cite{pete}. Hurst was a hydrologist who worked on the Nile River Dam project in the first decades of this century. At that time, it was common to assume that the uncontrollable influx of water from rainfall followed a random walk, in which each step is random and independent from previous ones. The random walk is based on the fundamental concept of Brownian motion. Brownian motion refers to the erratic displacements of small solid particles suspended in a liquid. The botanist Robert Brown, about 1828, realized that the motion of the particles is due to light collisions with the molecules of the liquid.

Hurst measured how the reservoir level fluctuated around its average level over time. The range of this fluctuation depends on the length of time used for measurement. If the series were produced by a random walk, the range would increase with the square root of time as $T^{1/2}$. Hurst found that the random walk assumption was wrong for the fluctuations of the reservoir level as well as for most natural phenomena, like temperatures, rainfall and sunspots. The fluctuations for all this phenomena may be characterized as a ``biased random walk''-a trend with noise- with range increasing as $T^H$, with $H > 0.5$. Mandelbrot called this kind of generalized random walk {\sl fractional brownian motion}. In high-school statistical courses we have been taught that nature follows the gaussian distribution which corresponds to random walk and $H=1/2$. Hurst's findings show that it is wrong.

The proper range for $H$ is from $0$, corresponding to very rough random fractal curves, to 1 corresponding to rather smooth looking fractals. In fact, there is a relation between $H$ and the fractal dimension $D$ of the graph of a random fractal: 
\begin{equation}
D=2-H.
\end{equation} 
Thus, when the exponent $H$ vary from $0$ to $1$, yields dimensions $D$ decreasing from $2$ to $1$, which correspond to more or less wiggly lines drawn in two dimensions.

Fractional Brownian motion can be divided into three distinct categories: $H<1/2$, $H=1/2$ and $H>1/2$. The case $H=1/2$ is the ordinary random walk or Brownian motion with independent increments which correspond to normal distribution.

For $H<1/2$ there is a negative correlation between the increments. This type of system is called {\it antipersistent}. If the system has been up in some period, it is more likely to be down in the next period. Conversely, if it was down before, it is more likely to be up next. The antipersistence strength depends on how far $H$ is from $1/2$.

For $H>1/2$ there is a positive correlation between the increments. 
This is a {\it persistent} series. If the system has been up (down) 
in the last period, it will likely continue positive (negative) in the 
next period. Trends are characteristics of persistent series. The strength 
of persistence increases as $H$ approaches $1$. Persistent time series are 
plentiful in nature and in social systems. As an example the Hurst exponent 
of the Nile river is $0.9$, a long range pesistence that requires unusually 
high barriers, such as the Asw\^an High Dam to contain damage in the floods.

 \section{Multifractals}

It can be said that a set's defining relation is an {\sl indicator function} associated to a point which can only take two values: {\sl true} or $1$ if the point belongs to the set; and {\sl false} or $0$ if the point does not belong to the set. However, most facts about nature demand more general mathematical objects to embody the idea of {\sl shades of grey}. Those objects are called {\sl measures}.

A simple example of multifractal may be obtained by considering a map of a continent. A possible measure $\mu$ is {\sl the number of people}. To each subset $S$ of the map, the measure lays a quantity $\mu(S)$, which is the number of people on $S$.  If we divide the map into two equal size parts $S_1$ and
$S_2$, $\mu(S_1)$ and $\mu(S_2)$ respectfully will be different. The
division can be done several times giving $\mu(S_i)$. Some countries have more people than others $\rightarrow$ parts of a country contain more people than others and so on. $ \mu =${\sl number of people} is a measure irregular at many scales.  When the irregularity is (at least statistically) the same at all scales, the measure is {\sl self similar} or {\sl multifractal}. 

For Euclidean support of a self-similar measure,  the box counting dimension  only confirms that there is nothing fractal about this support. Thus  $D$ gives not enough quantitative description
about the self similar measure supported by this set. What we are seeking is a measure given by a
weight which can be thought as the average density of probability in each
box, defined as $\mu(S)/{\epsilon}^E$ in a Euclidean space of dimension $E$ (or in a space of embedding dimension $E$).

For fractals,  instead of density, one speaks in terms of the {\sl coarse H\"older exponent $\alpha$}:
\begin{equation}
\alpha = \frac{log \mu(box)}{log \epsilon}.
\end{equation}

For a multifractal $\alpha$ will be restricted to an interval
$\alpha_{min}<\alpha<\alpha_{max}$ while for a fractal there will be an
unique $\alpha$.
To obtain the frequency distribution $f(\alpha)$, one must count the
number $N(\epsilon)$ of boxes of size $\epsilon$ that have a coarse H\"older
exponent $\alpha$.  Now suppose that a box of side $\epsilon$ has been selected at random among boxes whose total number is proportional to
 $\epsilon^{-E}$.  The probability of hitting
$\alpha$ is $p_{\epsilon}(\alpha) = N_{\epsilon}(\alpha)/{\epsilon}^{-E}$. In the case of interest to us, this distribution no longer tends to a limit as, $\epsilon \rightarrow 0$. Thus, instead of $p_{(\epsilon)}$ we use 
$f_{\epsilon}(\alpha)= -\frac{log N_{\epsilon}(\alpha)}{log \epsilon^{-E}}$.  As
$\epsilon \to 0$,  $\alpha$ becomes the {\it singularity exponent} and 
$f_{\epsilon}(\alpha)$ tends to the {\it singularity spectrum} $f(\alpha)$.
This implies that for each $\alpha$ the number of boxes increases for
decreasing $\epsilon$ as $N_{(\epsilon)}(\alpha) \sim 
{\epsilon}^{f(\alpha)}$.   $f(\alpha)$ is an upsidedown bell shaped curve,
which values could be interpreted as a fractal dimension of the subsets of boxes of
size $\epsilon$.  When  $\epsilon \to 0$  there are infinite subsets, each
characterized by its own $\alpha$ and a fractal dimension $f(\alpha)$.

\section{The multifractal formalism}
The aim of this formalism is to determinate the f($\alpha$) singularity
spectrum of a measure $\mu$ .
A partition function $Z$ can be defined from this spectrum (it is the same
model as the thermodynamic one).

\begin{equation}
Z(q,\epsilon) = \sum_{i=1}^{N(\epsilon)}\mu_i^q(\epsilon) \sim
\epsilon^{\tau(q)}  \:\:\:\:for \: \epsilon  \to 0.
\end{equation}

Both functions, $f(\alpha)$ and $\tau(q)$, describe the same aspects of a 
multifractal, and they are related to each other. In fact, the relationships 
are

\begin{equation}
 \tau(q) = f(\alpha) - q \alpha,
\end{equation}

where $\alpha$ is given as a function of $q$ by the solution of the equation 

\begin{equation}
 \frac{d}{d \alpha}(q \alpha - f(\alpha)) = 0.
\end{equation}
 
These two equations represent a {\it Legendre transform} from the variables $q$ 
and $\tau$ to the variables $\alpha$ and $f$.

The spectrum of $generalized\: fractal\: dimensions\: D_q$ is obtained from
the spectrum $\tau(q)$

\begin{equation}
D_q = \frac{\tau(q)}{(q-1)},
\end{equation}

The capacity or box dimension of the support of the distribution is given
by $D_0=f(\alpha(0))= -\tau(0)$.

$D_1=f(\alpha(1))=\alpha(1)$ corresponds to the scaling behavior of the
information and is called {\it information  dimension}. $D_1$ plays an 
important role in the analysis of nonlinear dynamic systems, especially in 
describing the loss of information as a chaotic system evolves in time.

For $q\ge2$, $D_q$ and the {\it q-point correlation integrals} are related. 
$D_2$ is called {\it correlation dimension} because it is associated with the 
``correlation function'' of the fractal set, that is, the probability of 
finding, within a distance of a given member of the set, another member 
\cite{gras}. 

As we will show in the following section the wavelet transform is especially 
suited to analyze a time series as a multifractal. 
 
\section{Wavelet Transform WT}
The work of Jean Morlet, a geophysicist with the oil company Elf-Aquitaine, who developed wavelets as a tool for oil prospecting, is usually taken as the starting point of the history of wavelets.

The standard way to look for underground oil is to send vibrations under ground and to analyze their echos to obtain the deepness and thickness of the layers and what materials they are made of. The problem is that there are hundreds of layers and all the signals interfere with each other. Fourier analysis was used to get information from the interference of the echos. As more powerful were the computers available, more Fourier windows were placed here and there. But the finer local definition needed to have access to information on different thicknesses layers couldn't be achieved.

In windowed Fourier analysis, a small window is ``blind'' to low frequencies, which correspond to signals too large for the window. On the other hand, large windows lose information about a brief change. Instead of keeping the size of the window fixed and change the frequencies of oscillations that filled the window, Morlet did the reverse: he kept constant the number of oscillations in the window and stretched or compressed the window like an accordion. This makes it possible to analyze a signal at different scales. The {\it wavelet transform (WT)} is sometimes called a ``mathematical microscope'': big wavelets give an approximate image of the signal, while smaller wavelets zoom in on details \cite{hubbard}. 

The wavelet transform of a signal $s(t)$ consists in decomposing it into
frequency and time coefficients, asociated to the wavelets.  The analyzing
wavelet $\psi$, by means of translations and dilations, generates the so
called family of wavelets.

The wavelet transform turns the signal $s(t)$ into a function
$T_\psi[s](a,b)$:

\begin{equation}
T_\psi[s](a,b)= \frac{1}{a}\int {\psi}^{*}[\frac{t-b}{a}]s(t)dt,
\end{equation}
where ${\psi}^{*}$ is the complex conjugate of $\psi$, $a$ is the frequency
dilation factor and $b$, the time translation parameter.

The wavelet to apply must be chosen with the condition:
\begin{equation}
\int  \psi(t)dt=0,
\end{equation}

and to be orthogonal to lower order polynoms
\begin{equation}
\int  t^m\psi(t)dt=0 \:\:\:\:\:\:\: 0\le m\le n;
\end{equation}  
where $m$ is the order of the polynom.

  In other words, lower order
polynomial behavior is eliminated and we can detect and characterize
singularities even if they are masked by a smooth behavior. Eq (12) is usually called the {\it vanishing moments} property and determines what the wavelet ``doesn't see''. ``Wavelet analysis is a way of saying that one is sensitive to change'' says Yves Meyer, one of the fathers of wavelets \cite{meyer}. ``It's like our response to speed. The human body is only sensitive to accelerations, not to speed''. This characteristic enables wavelets to compress information and, most important for us, makes them specially suited to study rough shapes like fractals or multifractals or to detect
 self-similarity or
self-affinity in time series. It has been used to study time series of completely different processes, like fractal scaling properties of DNA sequences or dissipation fields in fully developed turbulent flows, among many others \cite{arne}-\cite{turb}.\\
  For a value $b$ in the domain of the
signal, the modulus of the transform is maximized when the frequency $a$ is
of the same order of the characteristic frequency of the signal $s(t)$ in
the neighborhood of $b$, this last one will have a local singularity
exponent  $\alpha(b)\: \in \:]n,n+1[$. This means that around $b$ 

\begin{equation}
|s(t)-P_n(t)| \sim |t-b|^{\alpha(b)},
\end{equation}
where $P_n(t)$ is an $n$ order polynomial, and

\begin{equation}
T_{\psi}(a,b) \sim a^{\alpha(b)},
\end{equation}
provided the first $n+1$ moments are zero.  

If we have $\psi^{(N)}= d^{(N)}(e^{x^2/2})/dx^{N}$, the first $N$ moments
are vanishing. 

The Wavelet Modulus Function $|T_\psi[s](a,t)|$ will have a local maximum
around the points where the signal is singular.
These local maximum points make a geometric place called modulus maxima
line ${\cal L}$.

\begin{equation}
|T_\psi[s](a,b_l(a))| \sim a^\alpha(b_l(a))\:\:\:\:\: for \:\: a \to 0,
\end{equation}

where $b_l(a)$ is the position at the scale $a$ of the maximum belonging to
the the line ${\cal L}$.

The Wavelet Transform Modulus Maxima Method consists in the analysis of the
scaling behavior of some partition functions $Z(q,a)$ that can be defined as:

\begin{equation}
Z(q,a) = {\sum}|T_\psi[s](a,b_l(a))|^q,   
\end{equation}
and will scale like $a^{\tau(q)}$ \cite{arne}.
\begin{figure}[t]
\begin{center}
\psfig{figure=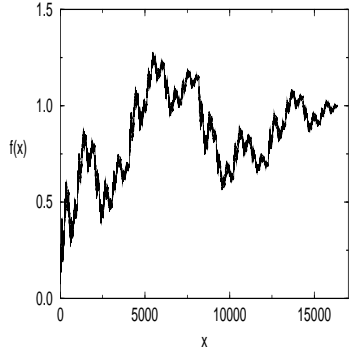,height=8cm,width=6cm}
\caption{\tt Generalized devil staircase}
\end{center}
\end{figure}

This partition function works like the previously defined partition function
for singular measures.  For $q>0$ will prevail the most pronounced modulus
maxima and, on the other hand, for $q<0$ will survive the lower ones.
The most pronounced modulus take place when very deep singularities are
detected, while the others correspond to smoother singularities.
We can get $\tau(q)$ (Eq. 3) and the Lagrange Transform can be applied,
obtaining $f(\alpha)$ and $D_q$ spectra, like in the previous section.
The shape of $f(\alpha)$ is a hump that has a maximum value for the
$\alpha^*$ associated with the general behavior of the series. So, this
particular singularity exponent can be thought like the Hurst exponent $H$
for the series.  

\section{Application of WTMM to a generalized \\ devil staircase}

\begin{figure}[t]
\begin{center}
\psfig{figure=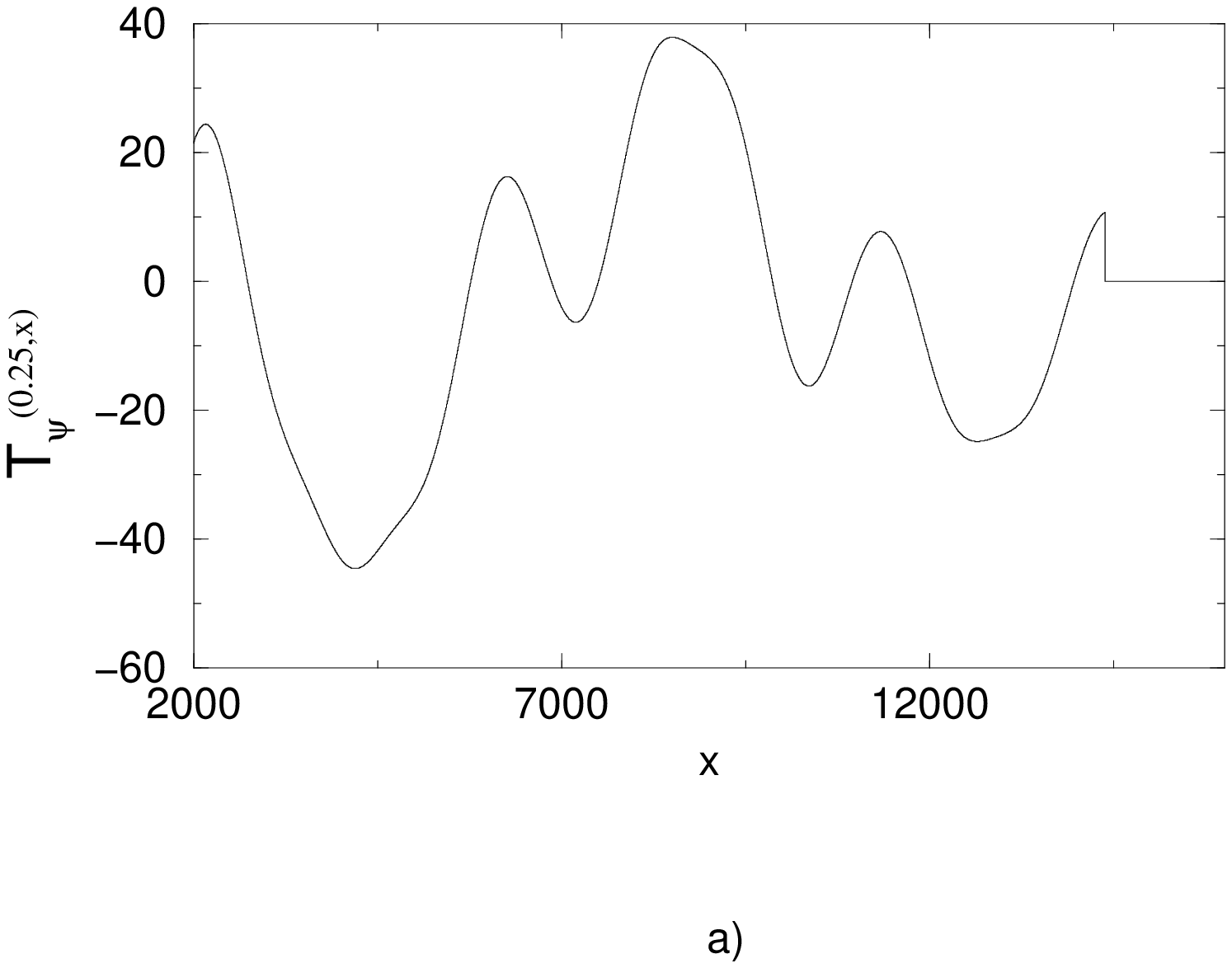,height=8cm,width=6cm}
\psfig{figure=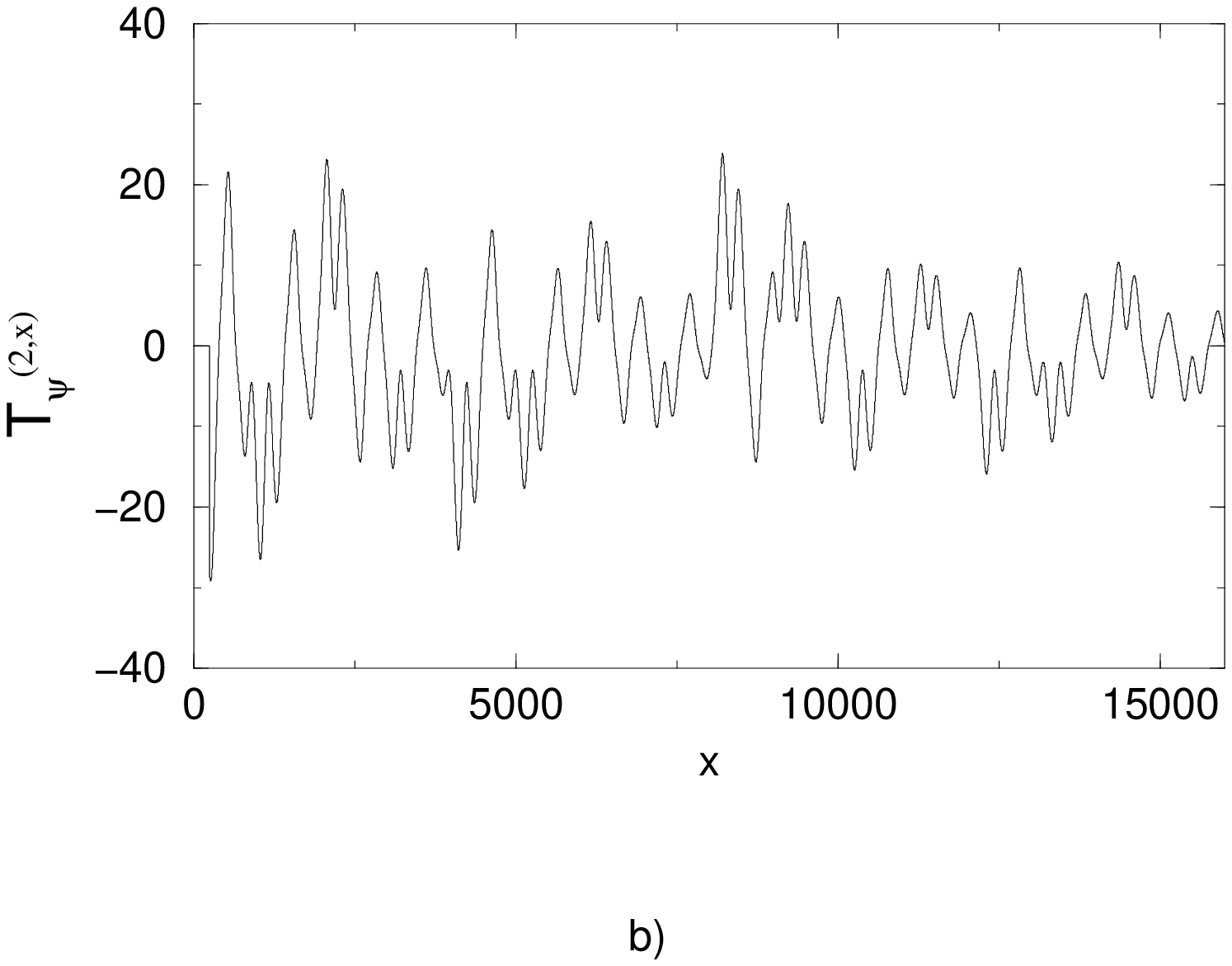,height=8cm,width=6cm}
\caption{\tt Generalized devil staircase: a) Wavelet transform with $a=4$. 
b) Wavelet transform with $a=0.5$.}

\end{center}
\end{figure}

Before applying the WTMM method to the analysis of social signals in time, it is
important to see how it works on "simple" functions, like self-similar measures
lying on "generalized" Cantor set. 

The {\it devil staircase} is a self-similar measure constructed recursively with
the Cantor set, but giving different weights or probabilities to the different
segments at each step (see \cite{hein}). The devil staircase is the distribution
function associated with the final probabilities.
    
Let $f$ be  a generalized devil staircase constructed recursively as follows:
each interval at each step of the construction is divided into four subintervals
of the same length on which we distribute respectively the weights $p_1=0.69$,
$p_2=-p_3=0.46$ and $p_4=0.31$. 

Fig. 4 displays $f(x)$. Two wavelet transforms of different scaling factor $a$
are shown in Fig. 5. $\tau(q)$ and $f(\alpha$) singularity spectra are displayed
in Fig. 6 a) and b). This generalized devil staircase is thus an everywhere 
singular signal
that displays multifractal properties.

\begin{figure}[t]
\begin{center}
\psfig{figure=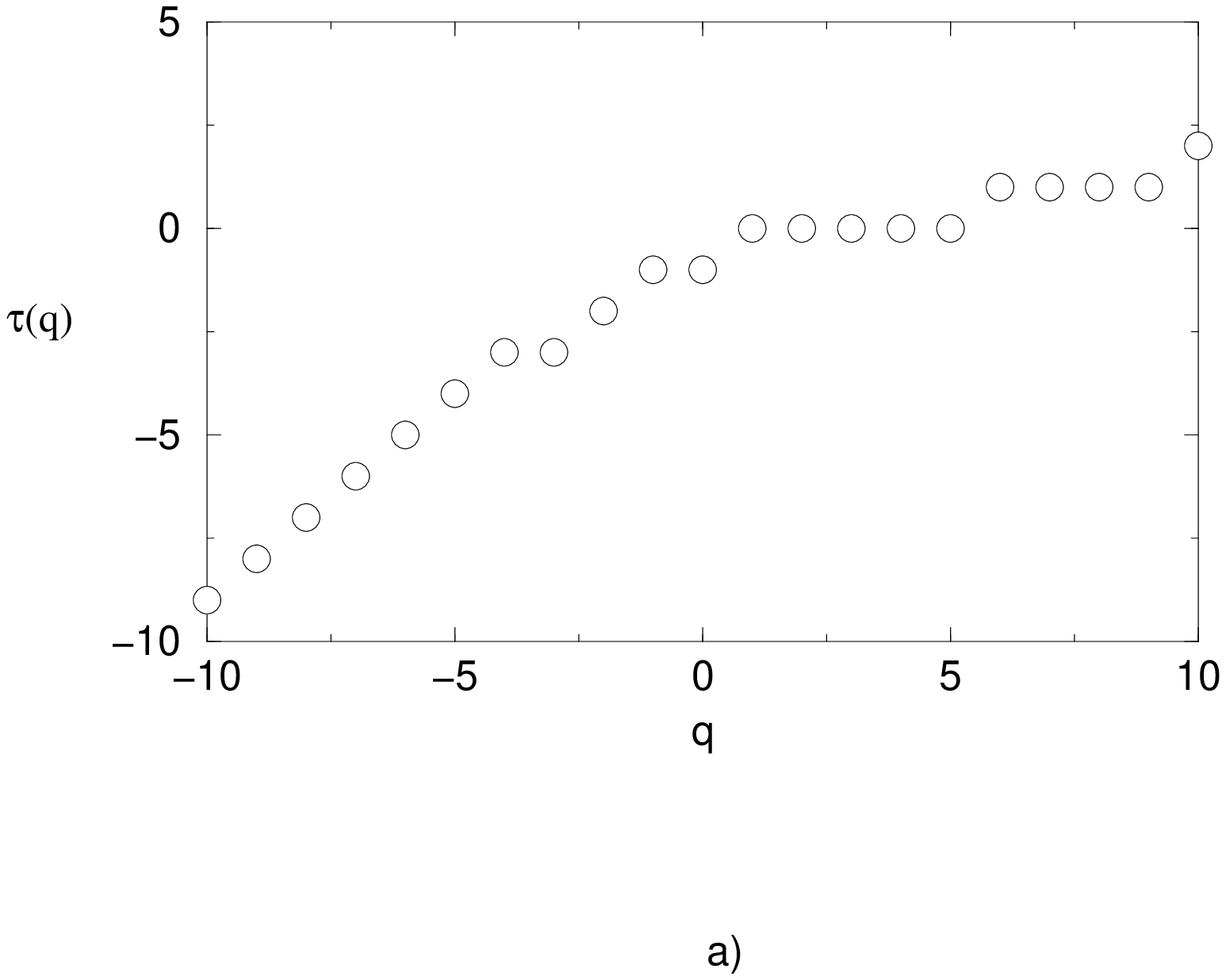,height=8cm,width=6cm}
\psfig{figure=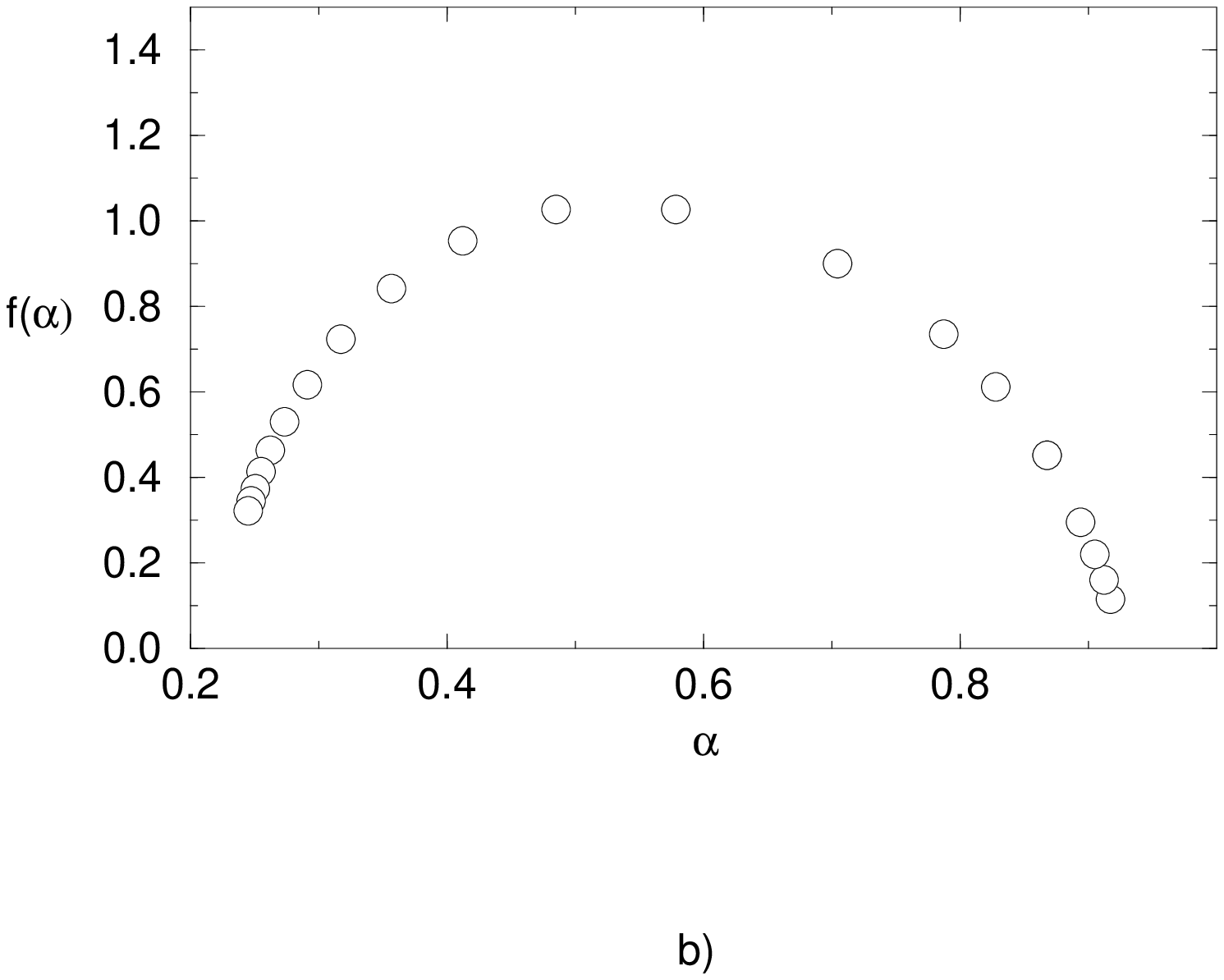,height=8cm,width=6cm}
\caption{\tt Generalized devil staircase spectrums:
a) $\tau(q)$ versus $q$. b) $f(\alpha)$ versus $\alpha$.}
\end{center}
\end{figure}As we mentioned above, persistent processes are common in nature.  Applying the
WTMM we will be able to verify that the electrical demand is a persistent
time series and to compare the quality of the process in two different
places. 
 
\section{Application of WTMM to electrical \\ demand  time series}

As an example of application of wavelets to a fractal time series, we chose 
the electrical demand of two completely different places: Australia, a whole 
continent, and Mar del Plata a touristic city of Argentina.\\
Data of Australia electrical power demand were obtained  at the web site:\\
http://www.tg.nsw.gov.au/sem/statistics.\\
Mar del Plata electrical demand time series was kindly
provided by Centro Operativo de Distribucion Mar del Plata belonging to
EDEA. 

Two time series of 8832 points were taken, as seen in Fig. 7 a) and b).

\begin{figure}[t]
\begin{center}
\psfig{figure=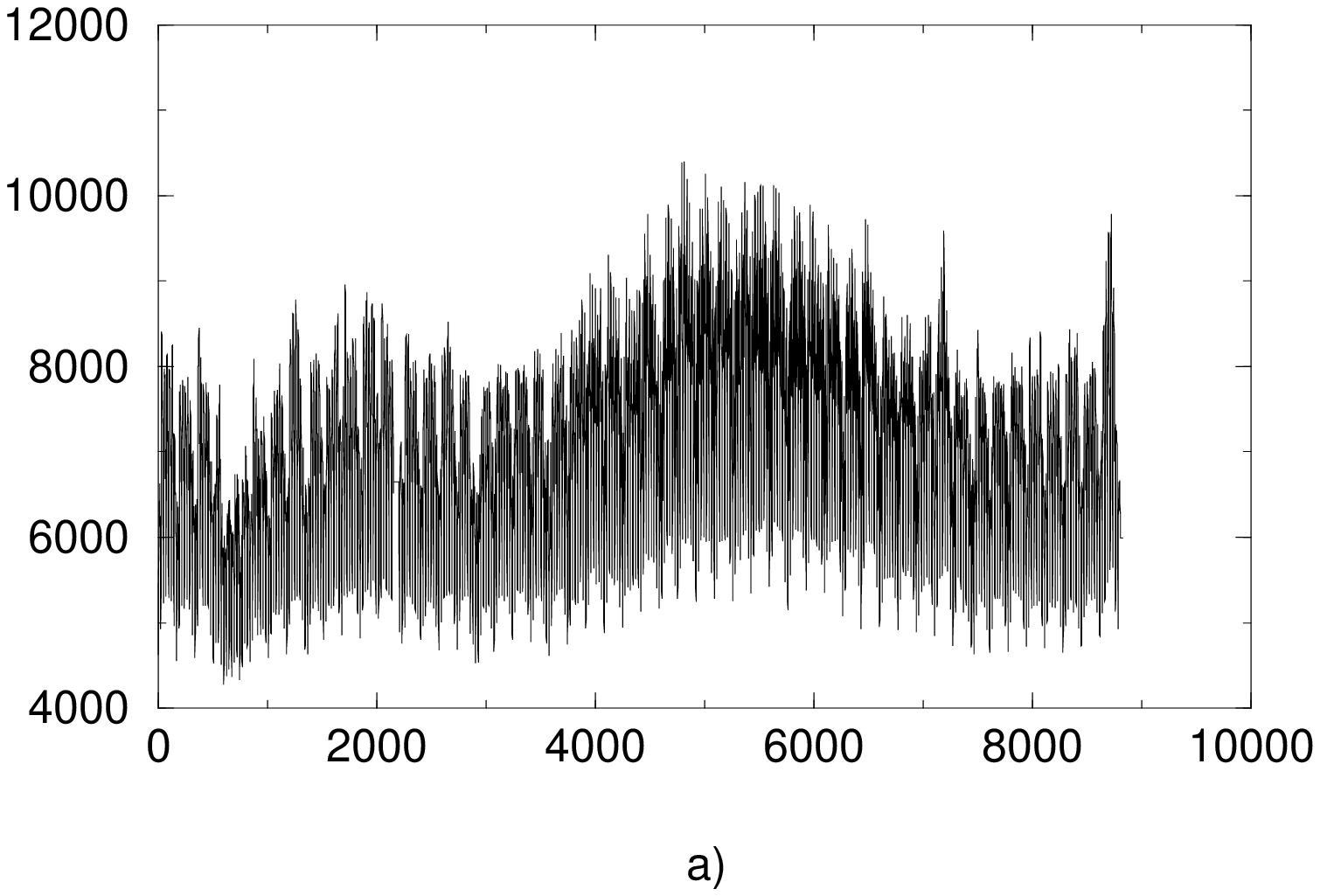,height=8cm,width=6cm}
\psfig{figure=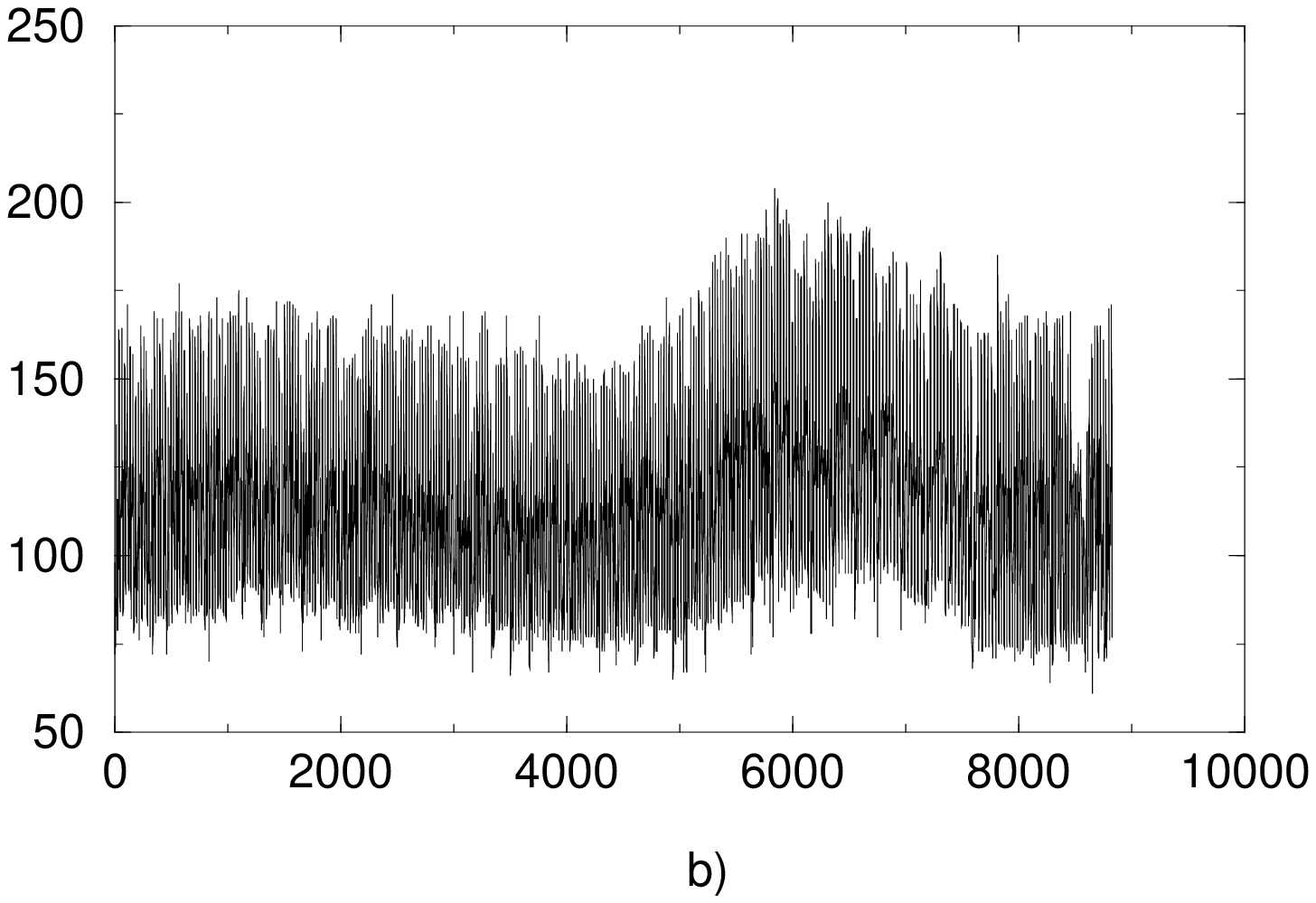,height=8cm,width=6cm}
\caption{\tt a) Electrical power demand time series for Australia. b) Electrical power demand time series for Mar del Plata.}
\end{center}
\end{figure}

The
fifth derivative of Gaussian function was chosen as analyzing wavelet:

\begin{equation}
\psi^{(5)}(t)=d^{(5)}(e^{t^2/2})/dt^{5},
\end{equation}

Twelve wavelet transform data files were obtained applying the Wavelet
Transform to both electrical demand data with $\psi^{(5)}$, ranging the
scaling factor $a$ from
$a_{min}=1/256$ to $a_{max}=8$ in steps of a power of two. 

Then we computed the partition function for each one for $-18\le q\le30$,
getting $\tau(q)$, as shown in Fig. 8 a) and b).

\begin{figure}[t]
\begin{center}
\psfig{figure=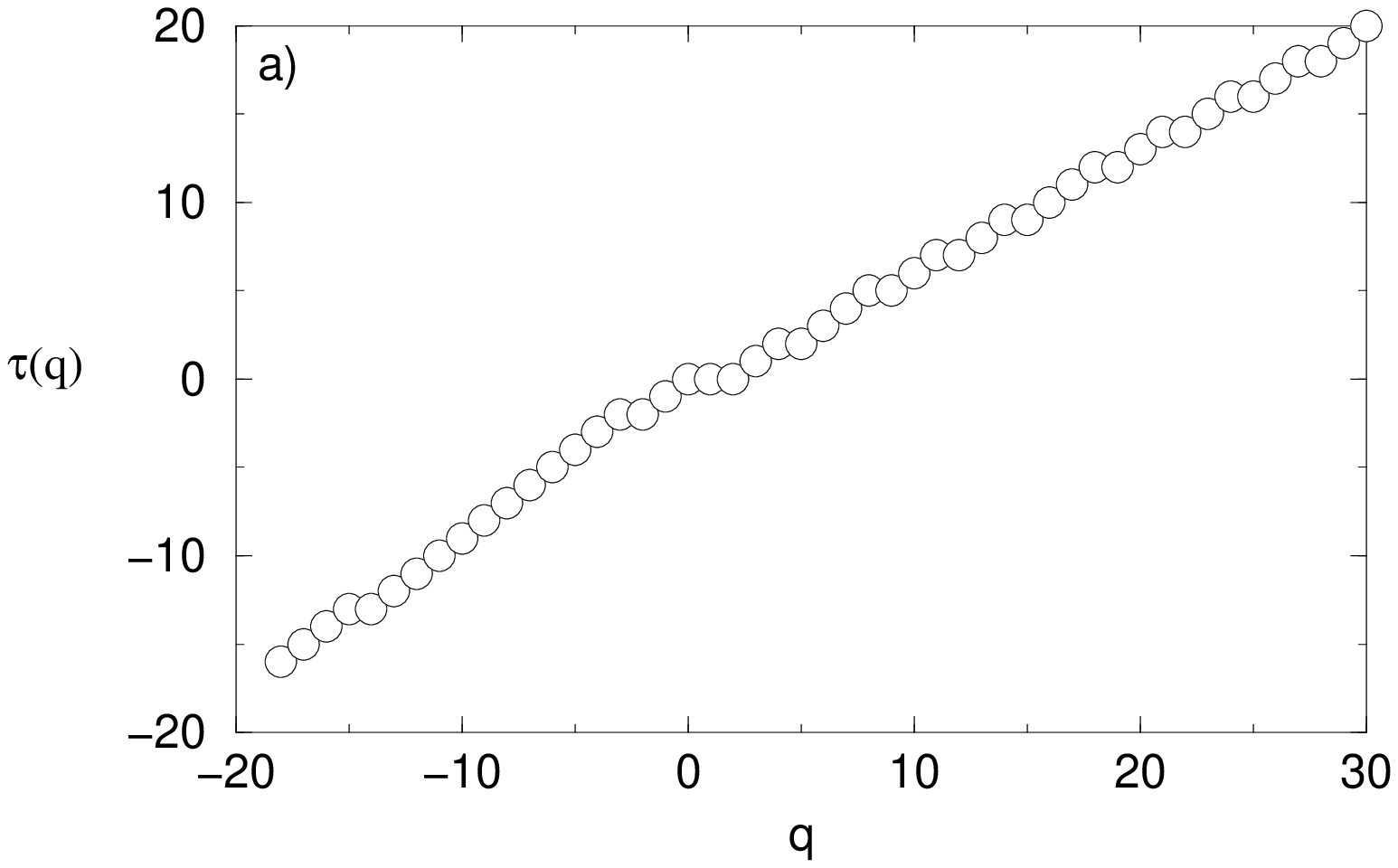,height=8cm,width=6cm}
\psfig{figure=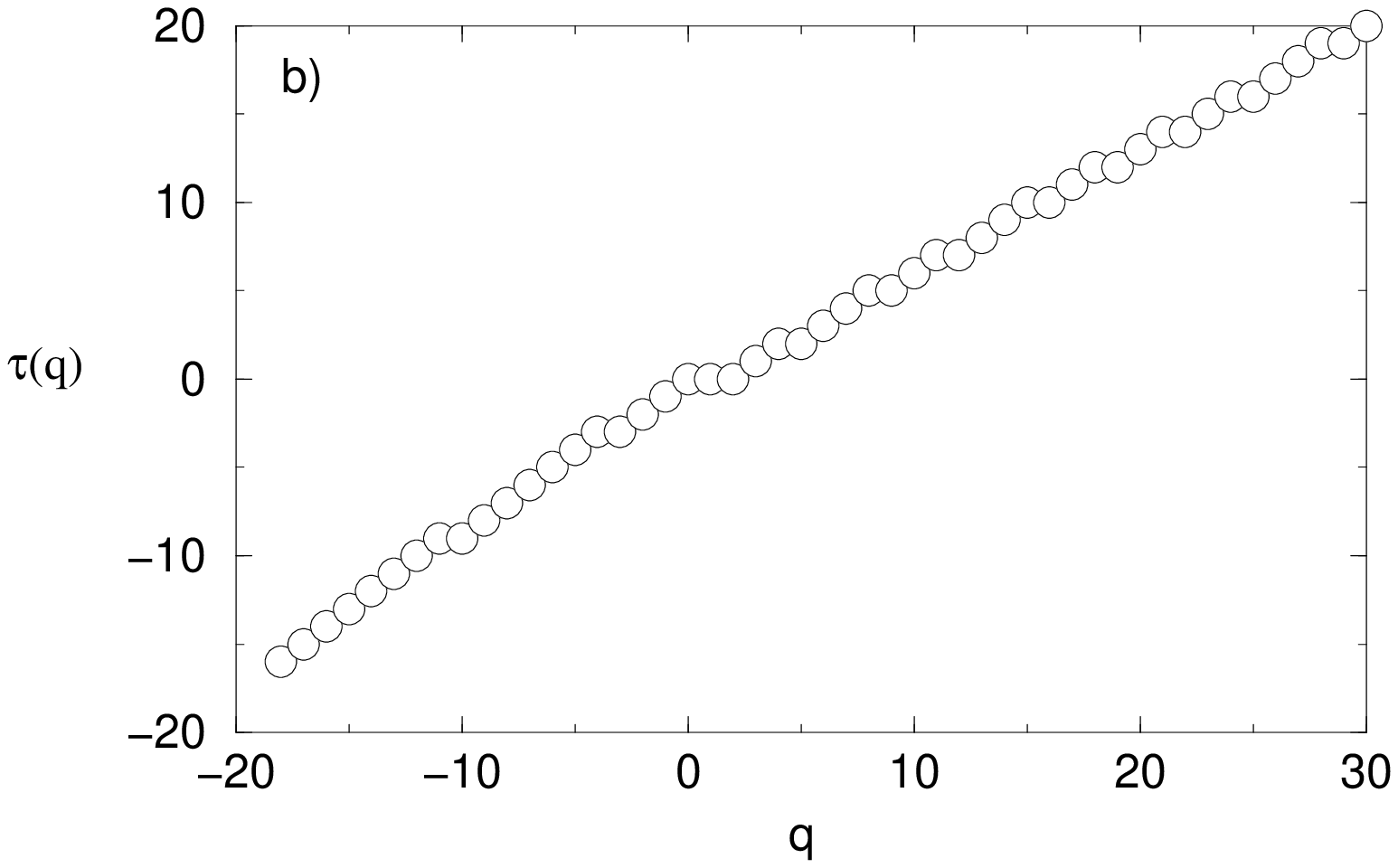,height=8cm,width=6cm}
\caption{\tt a) $\tau(q)$ for Australia electrical power demand. b)  $\tau(q)$ for Mar del Plata electrical power demand.}
\end{center}
\end{figure}

\begin{figure}[t]
\begin{center}
\psfig{figure=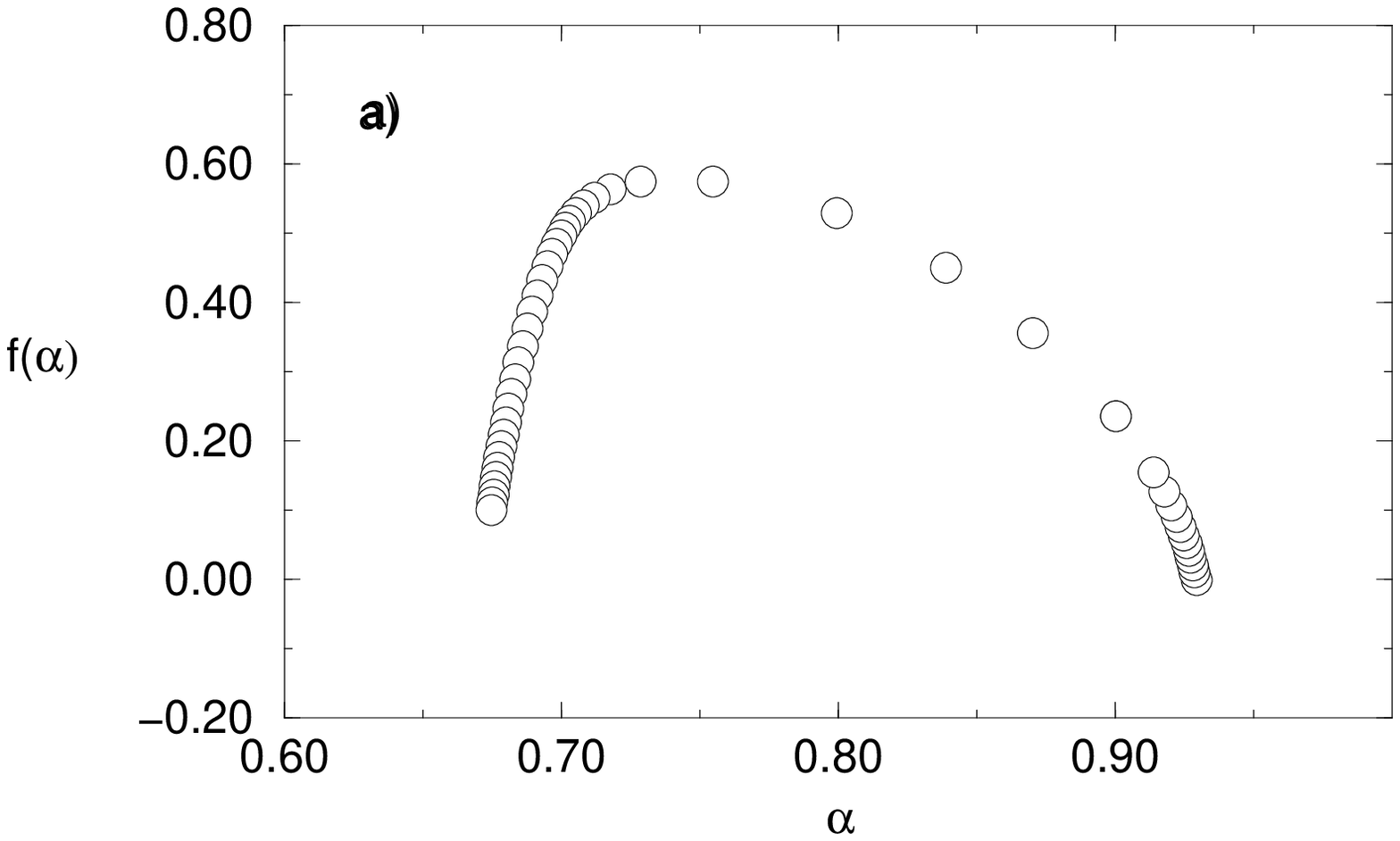,height=8cm,width=6cm}
\psfig{figure=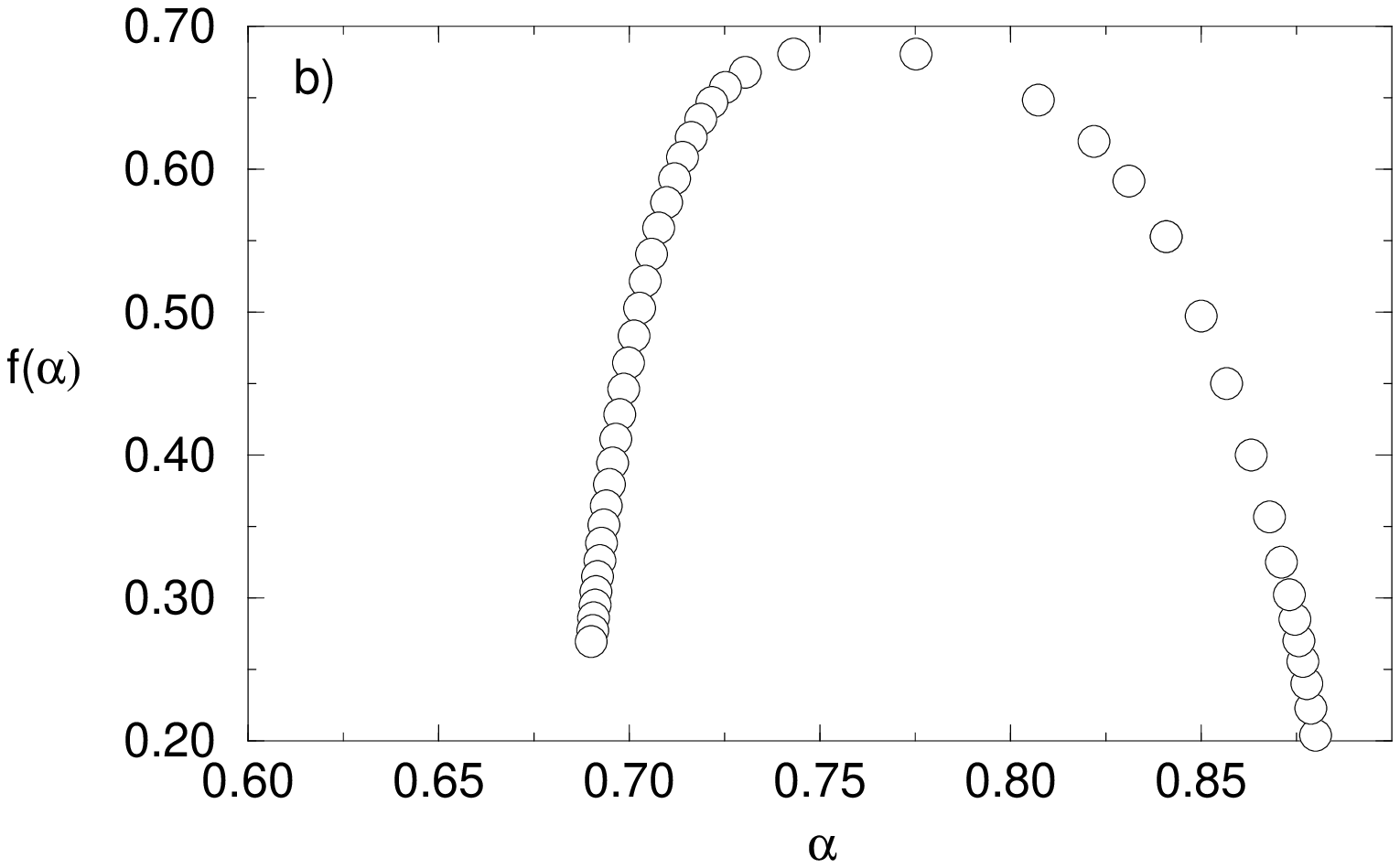,height=8cm,width=6cm}
\caption{\tt a) $f(\alpha)$ for Australia electrical power demand. b)  $f(\alpha)$ for Mar del Plata electrical power demand.}
\end{center}
\end{figure}

$\tau(q)$ is a nonlinear convex increasing function.  For Australia, 
$\tau(0)=-0.68$ and two slopes which are $\alpha_{min}=0.70$ for $q\le0$
and $\alpha_{max}=0.87$ for $q>0$, while for Mar del Plata city
$\tau(0)=-0.57$, $\alpha_{min}=0.69$ and $\alpha_{max}=0.92$ .

The corresponding $f(\alpha)$ singularity spectra obtained by
Legendre transforming $\tau(q)$ are displayed in Fig. 9 a) and b).  A multifractal
signal is characterized by a single humped shape  with a nonunique
H\"older exponent like each of the graphs shown in Fig. 9.

As expected from $\tau(q)$, the support of $f(\alpha)$ extends over a
finite interval which bounds are $\alpha_{min}=0.70$ and
$\alpha_{max}=0.87$ for Australia, which is larger than the one for Mar del
Plata ranging from $\alpha_{min}=0.69$ and $\alpha_{max}=0.92$.

The minimum value, $\alpha_{min}$, corresponds to the strongest singularity
which characterizes the most rarified zone, whereas higher values exhibit
weaker singularities until $\alpha_{max}$ or weakest singularity which
corresponds to the densiest zone.  $\alpha_{min}$ and $\alpha_{max}$ both
between $0.5$ and $1$ correspond to a persistent process; although it can be
observed a very little less persistence for Australia than for Mar del
Plata due to the slighty shift of the curve to the right for the last one,
the processes are deeply persistent .   

The support dimensions $D_o=D_{max}=-\tau(0)$ are 0.68 and 0.57 for
Australia and Mar del Plata respectfully; which implies that the capacities
of the supports are fractional so we are in presence of two chaotic
processes.   

The H\"older exponent for the dimension supports, $\alpha(D_{max})$, are
0.74 (Australia) and 0.73 (Mar del Plata) . These particular $\alpha$
corresponds to $f(\alpha)_{max}$ or $D_{max}$ which implies that the events
with $\alpha=\alpha(D_{max})$ are the most frequent ones.  

$0.5<\alpha\le1 $ implies we are analyzing a persistent time series which
obeys to the ''Joseph Effect'' (In the Bible refers to 7 years of loom,
happiness and health and 7 years of hungry and illness).  This system has
long memory effects: what happens now will influence the future, so there
is a very deep dependence with the initial conditions.  It may be thought
like a Fractional Brownian Motion of $\alpha>0.5$.

A Hurst exponent of 0.73 or 0.74 describes a very persistent time series,
what is expected in a natural process involved in an inertial system. 
$\alpha$ can be known as H\"older Exponent or Singularity Exponent, too. 
If the distribution is homogeneous there is an unique $\alpha=H$ (for
example Fractional Brownian Motion), but if it is not there are several
exponents $\alpha$, like in these two cases.  The most frequent $\alpha$
will characterize the series and will play as Hurst exponent.

$\stackrel{-}{\alpha}=(\alpha_{min}+\alpha_{max})/2$ is almost the same for
Australia and Mar del Plata; in fact 0.79 for the first and 0.80 for the
second one; bigger in both cases to $\alpha(D_{max})$ (0.74 and 0.73
respectfully).  This implies that the curves are slightly humped to the
left, an effect that is more pronounced for the city than for the country
and a better precision is obtained for the $q>0$ branch, where the bigger
values will prevail (i.e. high changes in the demand which are more rare) 
The asymmetrical shape of the spectrum reveals more pronounced
inhomogeneities in the events associates with the $q<0$ branch, asociated
with the smaller values of power(i.e slight changes in demand which are
more ordinary).

$\alpha_{range}=(\alpha_{max}-\alpha_{min})$  is other indicator of the
behavior.  For Mar del Plata $\alpha_{range}$ is larger than for Australia.  
 
The information dimension for Australia is $D_1=f(\alpha(1))=f(0.70)=0.70$
which features the scaling behavior of the information while it is
$D_1=f(\alpha(1))=f(0.64)=0.64$ for Argentina.
$D_1$ is a fractional number in both cases. Then, in Australia and in Mar del Plata the electricity demand corresponds to chaotic systems with the problems of forecasting associated with them.

The correlation dimensions are $D_2=\tau(2)=0.79$ in the case of  Mar del Plata
while $D_2=\tau(2)=0.87$ for Australia.  The correlation
dimension characterizes a chaotic atractor and, besides, $D_2>1/2$
indicates the presence of long-range correlations. 

The long-range correlations are observed in some biological systems lacking
of a characteristic scale of time or length.  Such behavior may be
adaptative because the long-range correlations play the role of the
organizing principle for highly complex, non linear processes that generate
fluctuations on a wide range of time scales and, in adittion, the lack of
characteristic scale helps to prevent excessive mode locking that would
restrict the reaction of the organism.

As we can see, there is longer-range correlation for Australia,
implying  that for the case of an abrupt change of the demand Australia
electrical system will have a better answer.

\section{Conclusion}

We presented a brief introduction to fractals, multifractals and wavelets.
Since their birth, fractals have shown to be ubiquitous in nature, and, in the last years are finding their way in social and economic systems. As fractals are the geometry of chaos, and chaos is probably present in far from equilibrium processes such as social ones, they will be more frequently found in social systems in the near future. Wavelets are a specially useful mathematical tool to sudy fractals. As an example of its application we used the Wavelet Transform Modulus Maxima Method to compare the electrical power demands of 
 a touristic city, Mar del Plata, and a whole country, Australia. 
We found that both electrical demands behave, like most ones in nature, as  long term memory
phenomena.
In both cases, the fractal dimensions obtained correspond to chaotic processes.
In particular, the correlation dimensions found by this way tell us that
the series observed for Australia is longer-range correlated than the one
for Mar del Plata.  This lays that Australia power generating system is
better suited to satisfy oscilations in the demand. 
In spite of $\alpha$ ranges only within the $0.5<\alpha<1$ interval, the
greater value of Mar del Plata $\alpha range$ indicates that the demand
varies in a wider ranger, which features the variation in the demography.
We think that with these examples the reader can realize that the fractal 
analysis is especially suited to study the non-linear statistics of social 
systems.

\end{document}